\documentclass[iop]{emulateapj}
\usepackage{graphicx}
\usepackage{natbib}
\usepackage[colorlinks,urlcolor=blue,citecolor=blue,linkcolor=blue]{hyperref} 
\usepackage{xspace}
\usepackage{color}

\newcommand{\kms}     {~km~s$^{-1}$\xspace}
\newcommand{\mjy}     {~mJy~beam$^{-1}$\xspace}

\newcommand{\msun}    {~$M_{\sun}$\xspace}
\newcommand{\lsun}    {~$L_{\sun}$\xspace}
\newcommand{\cmd}   {~cm$^{-2}$\xspace}

\newcommand{\iras}   {IRAS~18162-2048\xspace}

\begin{document}

\title{Multiple monopolar outflows driven by massive protostars in IRAS~18162-2048}

\author{M. Fern\'andez-L\'opez\altaffilmark{1}}
\author{J.~M. Girart\altaffilmark{2}}
\author{S. Curiel\altaffilmark{3}}
\author{L.~A. Zapata\altaffilmark{4}}
\author{J.~P. Fonfr\'{\i}a\altaffilmark{3}}
\author{K. Qiu\altaffilmark{5}}

\altaffiltext{1}{Department of Astronomy, University of Illinois at Urbana--Champaign, 1002 West Green Street, Urbana, IL 61801, USA; manferna@illinois.edu}
\altaffiltext{2}{Institut de Ciencies de l'Espai, (CSIC-IEEC),Campus UAB, Facultat de Ciencies, Torre C5-parell 2, 08193 Bellaterra, Catalunya, Spain; girart@ieec.cat}
\altaffiltext{3}{Instituto de Astronom\'{\i}a, Universidad Nacional Aut\'onoma de M\'exico (UNAM), Apartado Postal 70-264, 04510 M\'exico, DF, Mexico}
\altaffiltext{4}{Centro de Radioastronom\'{\i}a y Astrof\'{\i}sica, UNAM, Apartado Postal 3-72, Morelia, Michoac\'an 58089, M\'exico}
\altaffiltext{5}{School of Astronomy and Space Science, Nanjing University, Nanjing 210093, China}


\begin{abstract}
In this paper we present Combined Array for Research in Millimeter-wave Astronomy (CARMA) 3.5~mm observations and SubMillimeter Array (SMA) 870~$\mu$m observations toward the high-mass star-forming region \iras, the core of the HH~80/81/80N system. Molecular emission from HCN, HCO$^+$ and SiO is tracing two molecular outflows (the so-called Northeast and Northwest outflows). These outflows have their origin in a region close to the position of MM2, a millimeter source known to harbor a couple protostars. We estimate for the first time the physical characteristics of these molecular outflows, which are similar to those of $10^3-5\times10^3$\lsun protostars, suggesting that MM2 harbors high-mass protostars. 
High-angular resolution CO observations show an additional outflow due southeast. We identify for the first time its driving source, MM2(E), and see evidence of precession. 
All three outflows have a monopolar appearance, but we link the NW and SE lobes, explaining their asymmetric shape as a consequence of possible deflection.
\end{abstract}

\section{Introduction}

\medskip

High-mass stars exert a huge influence on the interstellar medium, ejecting powerful winds and large amounts of ionizing photons or exploding as supernova. Despite their importance injecting energy and momentum into the gas of galaxies, how they form is still not well understood. High--mass protostars are typically more than 1~kpc from Earth (\citealt{2005Cesaroni}). They are deeply embedded inside dense molecular clouds and often accompanied by close members of clusters. Hence, well detailed studies have only been possible in a few cases. From a handful of studies based on high-angular resolution observations, we know that early B-type protostars posses accretion disks (e.g., \citealt{2001Shepherd,2005Cesaroni,2005Patel,2010GalvanMadrid,2010Kraus,2011FernandezLopez1}), while for O-type protostars there is no unambiguous evidence for accretion disks yet, but vast toroids of dust and molecular gas have been seen rotating around them, showing signs of infalling gas (e.g., \citealt{2005Sollins1,2005Sollins2,2006Beltran,2009Zapata,2012Qiu,2012JimenezSerra,2013Palau}). On the other hand, molecular outflow studies seem to display diverse scenarios in the few cases where both, high-angular resolution and relatively nearby targets were available.
For example, the two nearest to the Earth massive protostars (Orion BN/KL and Cep A HW2) have molecular outflows that are very difficult to interpret. It has been suggested that Orion BN/KL shows an  explosion-like isotropic ejection of molecular material \citep{2009Zapata}, while Cep A HW2 drives a very fast jet (\citealt{2006Curiel}), which is apparently pulsing and precessing due to the gravitational interaction of a small cluster of protostars \citep{2009Cunningham,2013Zapata}. These two examples provide how analyzing the outflow activity of these kind of regions can provide important insight on the nature of each region. They are also showing us that although accretion disks could be ubiquitously found around all kind of protostars, a complete understanding on the real nature of the massive star-formation processes should almost inevitably include the interaction between close-by protostars. This introduces much more complexity to the observations, not only because of the possible interaction between multiple outflows, but also the difficulty to resolve the multiple systems with most telescopes. 

\begin{deluxetable*}{clcccrrcc}[h]
\tablewidth{0pc}
\tablecolumns{10}
\tablecaption{\iras observations}
\tablehead{
\colhead{Tracks\tablenotemark{(a)}} & \colhead{Data} & \colhead{$\nu$\tablenotemark{(b)}} & \colhead{$\Delta$v}& \colhead{$\delta$v}& \colhead{$E_u$} & \multicolumn{2}{c}{Synthesized Beam}  & \colhead{RMS\tablenotemark{(c)}} \\
\colhead{} & \colhead{} & \colhead{(GHz)} & \colhead{(\kms)}& \colhead{(\kms)} & \colhead{(K)} & \colhead{$\arcsec\times\arcsec$} & \colhead{$\degr$} & \colhead{(mJy~beam$^{-1}$)} \\
}
\startdata
\multicolumn{9}{c}{CARMA observations} \\
\hline \\
3 & Continuum & 84.80200 &\nodata & \nodata & \nodata & $9.2\times5.8$ & 18 & 2 \\
3 & SiO (2-1)                                    & 86.84696  &400 & 5.2 &  6.3   & $8.6\times6.0$ & 7    & 20 \\                          
2 & HCO$^+$ (1-0)                          & 89.18852 &  400 & 5.2 & 4.3   & $9.0\times5.6$ & 14  & 30 \\
1 & HCN (1-0)                                   & 88.63160 &  400 & 5.2 &  4.4  & $9.3\times5.9$  & 14  & 30 \\      
1 & HC$_3$N (10-9)                          & 90.97902 &  400 & 5.2 & 24.0  & $8.9\times5.8$  & 17  & 35 \\
\hline \\
\multicolumn{9}{c}{SMA observations} \\
\hline \\
2 & CO (3-2)                                    & 345.79599  & 767 & 1.4 &  33.2   & $0.45\times0.34$ & 31    & 77 \\                          
1 & CO (3-2)                                    & 345.79599  & 357 & 1.4 &  33.2   & $3.0\times1.6$ & -6    & 117                       
\enddata 
\tablenotetext{(a)}{Number of tracks observed.}
\tablenotetext{(b)}{Data extracted from Cologne Database for Molecular Spectroscopy (CDMS) catalogue.}
\tablenotetext{(c)}{Per channel.}
\label{t_observ}
\end{deluxetable*}

\begin{figure}[h]
\epsscale{1}
\plotone{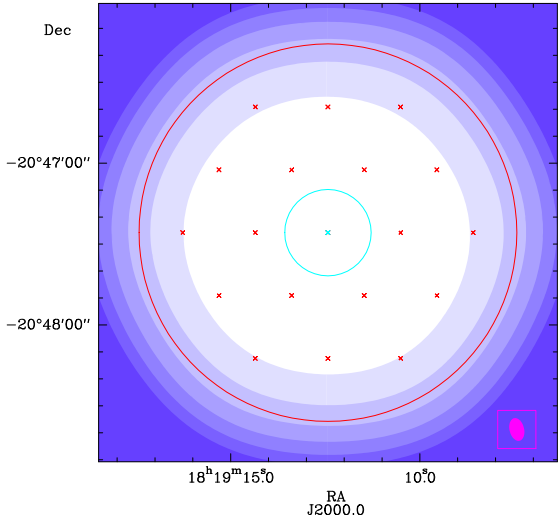}
\caption{The color scale image shows the sensitivity response of the CARMA mosaic pattern taken toward \iras. The white area is that with rms noise = 2\mjy, and each shaded zone has an rms noise of 1.1, 1.2, 1.3, 1.4 and 1.5$\times$2\mjy. The red crosses mark the center of each mosaic pointing and the red circle has a radius of 70$\arcsec$. Inside this red circle, the Nyquist-sampling of the mosaic insures an almost uniform rms noise. The synthesized beam is plotted in the bottom right corner.}
\label{contsen}
\end{figure}

\iras, also known as the GGD27 nebula, is associated  with the unique HH~80/81/80N radio jet. It is among the nearest high--mass protostars (1.7~kpc). The large radio jet ends at HH~80/81 in the south \citep{1993Marti} and at HH~80N in the north \citep{2004Girart}, wiggling in a precessing motion across a projected distance of 7.5~pc \citep{1998Heathcote}. Recently, \cite{2012Masque} have proposed that this radio jet could be larger, extending to the north, with a total length of 18.4~pc. Linear polarization has been detected for the first time in radio emission from this jet indicating the presence of magnetic fields in a protostellar jet \citep{2010CarrascoGonzalez}. The jet is apparently launched close to the position of a young and massive protostar, surrounded by a very massive disk of dust and gas rotating around it 
\citep{2011FernandezLopez1,2011FernandezLopez2,2012CarrascoGonzalez}. 
In spite of this apparently easy-to-interpret scenario, the region contains other massive protostars \citep{2009Qiu, 2011FernandezLopez1} which are driving other high velocity outflows.
(Sub)mm observations of the central part of the radio jet revealed two main sources, MM1 and MM2, separated about 7$\arcsec$ \citep{2003Gomez}, that are probably in different evolutionary stages \citep{2011FernandezLopez1}. MM1 is at the origin of the thermal radio jet, while MM2 is spatially coincident with a water maser and has been resolved into a possible massive Class 0 protostar and a second even younger source \citep{2011FernandezLopez2}. MM2 is also associated with a young monopolar southeast CO outflow \citep{2009Qiu}. There is evidence of a possible third source, MC, which is observed as a compact molecular core, detected only through several millimeter molecular lines. At present, MC is interpreted as a hot core, but the observed molecular line emission could also be explained by shocks originated by the interaction between the outflowing gas from MM2 and a molecular core or clump \citep{2011FernandezLopez2}. 

In this paper we focus our attention on the CO(3-2), SiO(2-1), HCO$^+$(1-0) and HCN(1-0) emission toward \iras, aiming at detecting and characterizing outflows from the massive protostars inside the \iras core and their interaction with the molecular cloud. Millimeter continuum and line observations of \iras were made with CARMA, while submillimeter continuum and line observations were made with the SMA. In Section 2, we describe the observations undertook in this study. In Section 3 we present the main results and in section 4 we carry out some analysis of the data. Section 5 is dedicated to discuss the data, and in Section 6 we give the main conclusions of this study.

\section{Observations}
\subsection{CARMA}
\iras observations used the CARMA 23-element mode: six 10.4~m antennas, nine 6.1~m antennas, and eight 3.5~m antennas. This CARMA23 observing mode included up to 253 baselines that provide extra short spacing baselines, thus minimizing missing flux when observing extended objects. 

The 3.5~mm (84.8~GHz) observations were obtained in 2011 October 9, 11 and 2011 November 11. The weather was good for 3~mm observations in all three nights, with $\tau_{230GHz}$ fluctuating between 0.2 and 0.6. The system temperature varied between 200 and 400~K. During those epochs, CARMA was in its D configuration and the baselines ranged from 5 to 148~m. The present observations are thus sensitive to structures between $6\arcsec$ and $70\arcsec$, approximately\footnote{Note that while the observations are often said to be {\it{sensitive}} to structures on scales $\Theta=\lambda/B_{min}\simeq206\arcsec~[\lambda(mm)/B_{min}(m)]$, the interferometer flux filtering could reduce the actual largest scale detection roughly in a factor 2, so that $\Theta\simeq100\arcsec~[\lambda(mm)/B_{min}(m)]$ (see appendix in \citealt{1994Wilner}). Here we report a value estimated using the last expression.}. We made a hexagonal nineteen pointing Nyquist--sampled mosaic with the central tile pointing at R.A.(J2000.0)=$18^h 19^m 12\fs430$ and DEC(J2000.0)=$-20\degr 47\arcmin 23\farcs80$ (see Fig. \ref{contsen}). This kind of mosaic pattern is used to reach a uniform rms noise in the whole area covered. The field of view of the mosaic has thus, a radius of about $70\arcsec$. Beyond this radius the rms noise increases over 20\%.

At the time of the observations, the correlator of the CARMA 23-element array provided 4 separate spectral bands of variable width. One spectral band was set with a bandwidth of 244~MHz, used for the 3.5~mm continuum emission. The other three bands were set with a bandwidth of 125~MHz, providing 80 channels with a spectral resolution of 1.56~MHz ($\sim5.2$\kms). One of these three bands was always tuned at the SiO (2-1) frequency. The other two bands were tuned at two of the following lines: HCO$^+$ (1-0), and HCN (1-0) and HC$_3$N (10-9).

The calibration scheme was the same during all the nights, with an on-source time of about 2 hours each. The gain calibrator was J1733-130 (with a measured flux density of 4.1$\pm$0.4~Jy), which we also used as the bandpass calibrator. Observations of MWC349, with an adopted flux of 1.245~Jy provided the absolute scale for the flux calibration. The uncertainty in the flux scale was 10\% between the three observations, while the absolute flux error for CARMA is estimated to be 15-20\%. In the remainder of the paper we will only consider statistical uncertainties.

The data were edited, calibrated, imaged and analyzed using the MIRIAD package (\citealt{1995Sault}) in a standard way. GILDAS\footnote{The GILDAS package is available at http://www.iram.fr/IRAMFR/GILDAS} was also used for imaging. Continuum emission was built in the uv--plane from the line-free channels and was imaged using a uniform weighting to improve the angular resolution of the data. For the line emission, the continuum was removed from the uv--plane. We then imaged the continuum-free emission using a natural weighting to improve the signal-to-noise ratio.  The continuum and the line images have rms noises of about 2\mjy and 20-35\mjy per channel, respectively.

\subsection{SMA}
The CO(3-2) maps at subarcsecond angular resolution presented in this paper were obtained from SMA observations taken on 2011 July 18 and October 3 in the extended and
very extended configurations, respectively.  The calibration, reduction and imaging procedures used in these observations are described in Girart et al. (2013, in preparation). Table 1 shows the imaging parameters for this data set.

The CO(3-2) maps at an angular resolution of $\simeq 2\arcsec$ presented in this paper were obtained from the public SMA data archive. The observations were done in the compact configuration on 2010 April 9. We downloaded the calibrated data and used the spectral windows around the CO(3-2) to produce integrated intensity maps.

\begin{figure*}
\epsscale{0.75}
\plotone{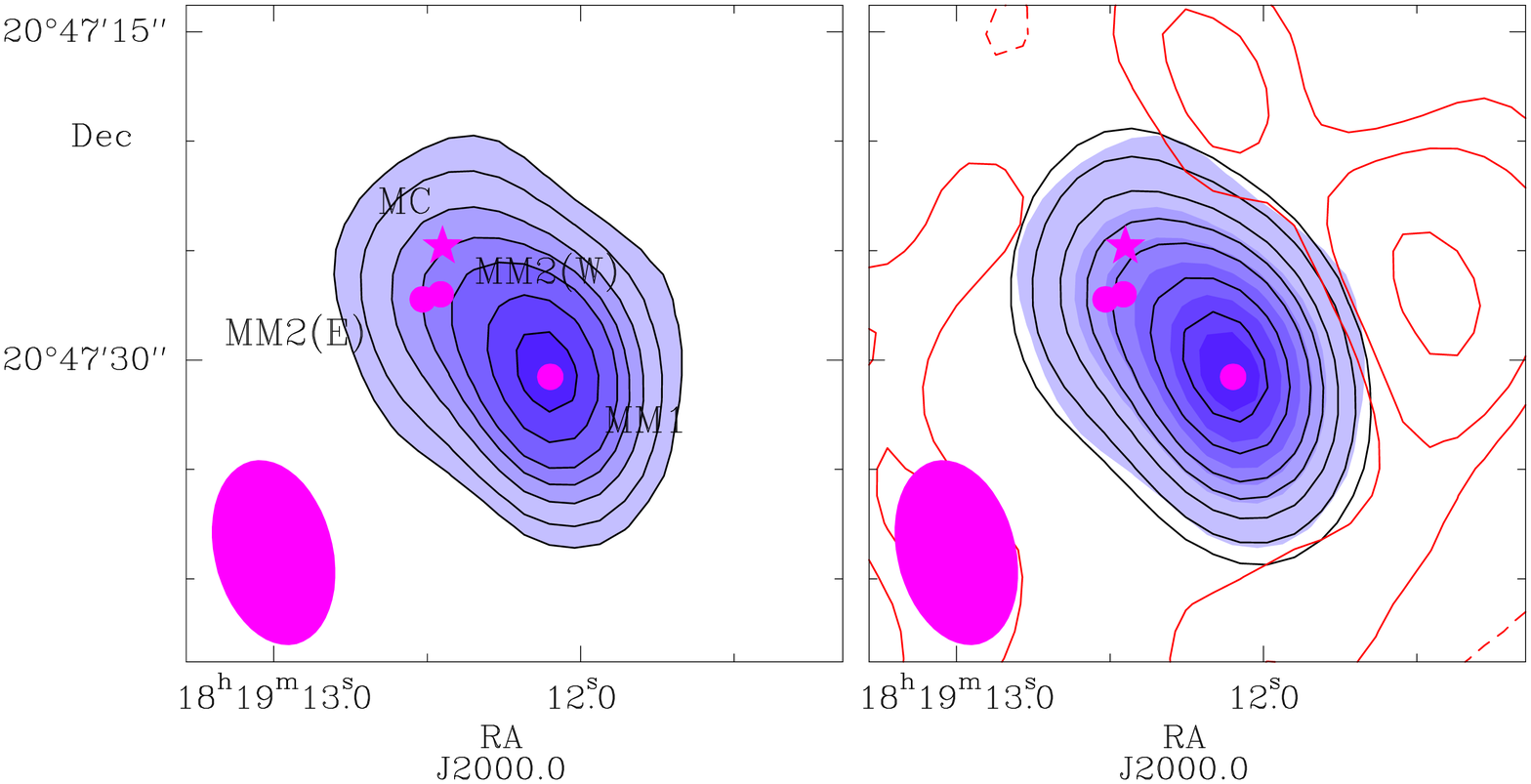}
\caption{{\bf Left:} CARMA image of the 3.5~mm continuum emission (black contours and color scale) toward the central region of \iras. Contours are 4, 6, 9, 12, 15, 18 and 21 $\times$ 2\mjy, the rms noise of the image. The star mark the position of the putative molecular core (\citealt{2009Qiu}) and the circles mark the position of the millimeter sources reported in \cite{2011FernandezLopez1}. The synthesized beam is shown in the bottom left corner. {\bf Right:} 3.5~mm continuum emission (color scale) overlapped with the two-Gaussian fit made to the image (black contours as in left panel, see \S 3.1) and the residuals left after the fit (red contours at -4, -3, -1.5, -0.75, 0.75, 1.5, 3 and 4 $\times$ 2\mjy).}
\label{cont}
\end{figure*}

\begin{figure*}
\epsscale{0.75}
\plotone{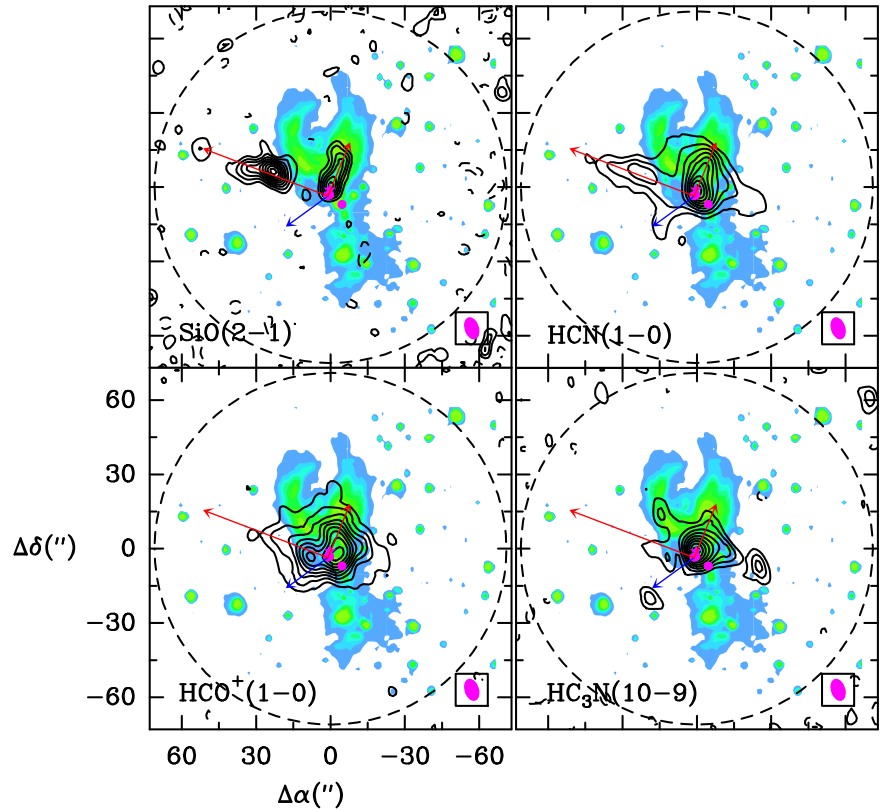}
\caption{Black contours represent the velocity-integrated flux maps of SiO(2-1) (top left panel), obtained by integrating the emission at about -1\kms and +31\kms; HCN(1-0) (top right panel), obtained by integrating the emission at about -7\kms and +47\kms; HCO$^+$(1-0) (bottom left panel) obtained by integrating the emission at about -12\kms and +10\kms; HC$_3$N(10-9) (bottom right panel), obtained by integrating the emission at about -1\kms and +4\kms. The contours are -35, -25, -15, 15, 25, 35, 45, 55, 65, 75, 85 and 95\% of the peak flux in all the maps. The color scale image represents the 2MASS K-band emission. The magenta circles and the star mark the position of the millimeter cores and the molecular core, respectively. Two red arrows show the redshifted NE and NW outflow directions, while the blue arrow show the blueshifted SE outflow path (see \S 3.2.1). The region inside the dashed circle has an almost uniform rms noise level. The synthesized beam is shown in the bottom right corner of each panel.}
\label{moms0}
\end{figure*}

\begin{figure}[h]
\epsscale{1}
\plotone{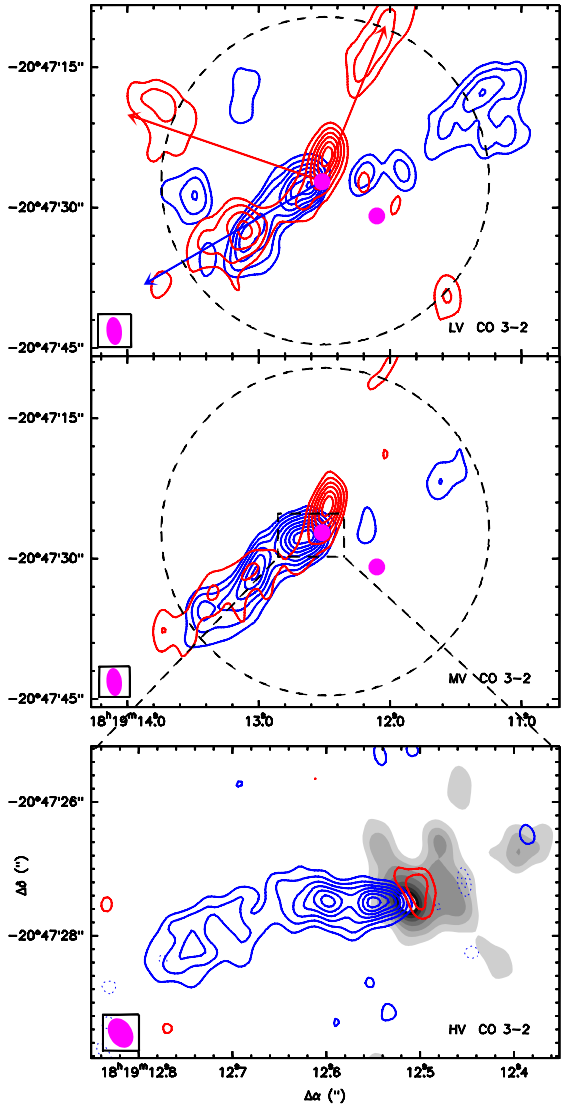}
\caption{SMA redshifted (red contours) and blueshifted (blue contours) emission of the CO(3-2) line. In the two upper panels the magenta circles mark the positions of MM1 and MM2. The dashed black circles represent the SMA primary beam and the synthesized beam is shown in the bottom left corner of each panel. {\bf Top panel:} LV component obtained by integrating the emission at about -15\kms and +20\kms (blue: from -22 to -10\kms; red: from +14 to +26\kms). Contour levels are 5, 13, 25, 40, 60, 80 and 100 $\times60$\mjy. The synthesized beam is $3\farcs0\times2\farcs6$, P.A.$=-6\degr$. {\bf Middle panel:} MV component obtained by integrating the emission at about -50\kms and +50\kms (blue: from -72 to -2\kms; red: from +24 to +74\kms). Contour levels are 5, 13, 25, 40, 60, 80 and 100 $\times10$\mjy. The synthesized beam is $3\farcs0\times2\farcs6$, P.A.$=-6\degr$. {\bf Bottom panel:} Zoom to the MM2 surroundings of the HV component obtained by integrating the emission at about -85\kms and +115\kms (blue: from -118 to -74\kms; red: from +106 to +130\kms). CO(3-2) contour levels are 3, 5, 7, 10, 14, 18, 22, 26 and 30 $\times3.2$\mjy. The synthesized beam is $0\farcs45\times0\farcs34$, P.A.$=31\degr$. The grey scale shows the SMA 870$\mu$m continuum emission from the easternmost MM2(E) and the westernmost MM2(W).}
\label{co32}
\end{figure}

\begin{figure*}
\epsscale{1}
\plotone{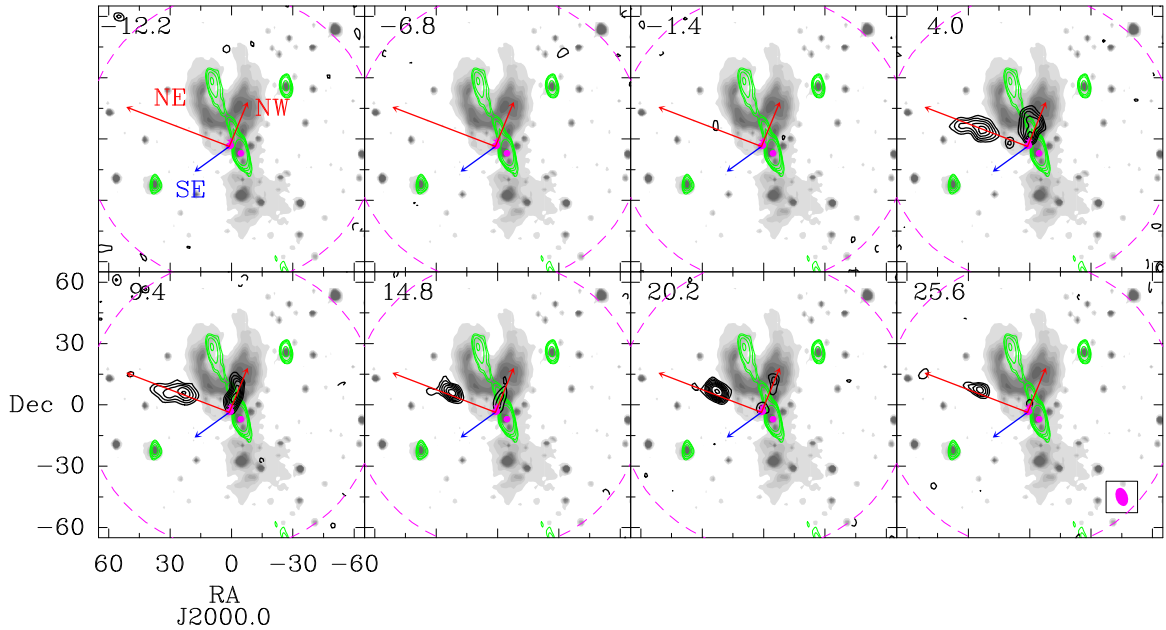}
\caption{SiO(2-1) emission velocity map (black contours) overlapped with a VLA HH~80/81/80N radio jet image (green contours) at 6~cm and a 2MASS K-band infrared image (grey scale). SiO(2-1) contours are -3, 3, 4, 5, 6, 7, 8, 9, 10, 12, 14 and 16 $\times$ 20\mjy, the rms noise of the image. Symbols are the same that those of Fig. \ref{moms0}.}
\label{sio}
\end{figure*}

\begin{figure*}
\epsscale{1}
\plotone{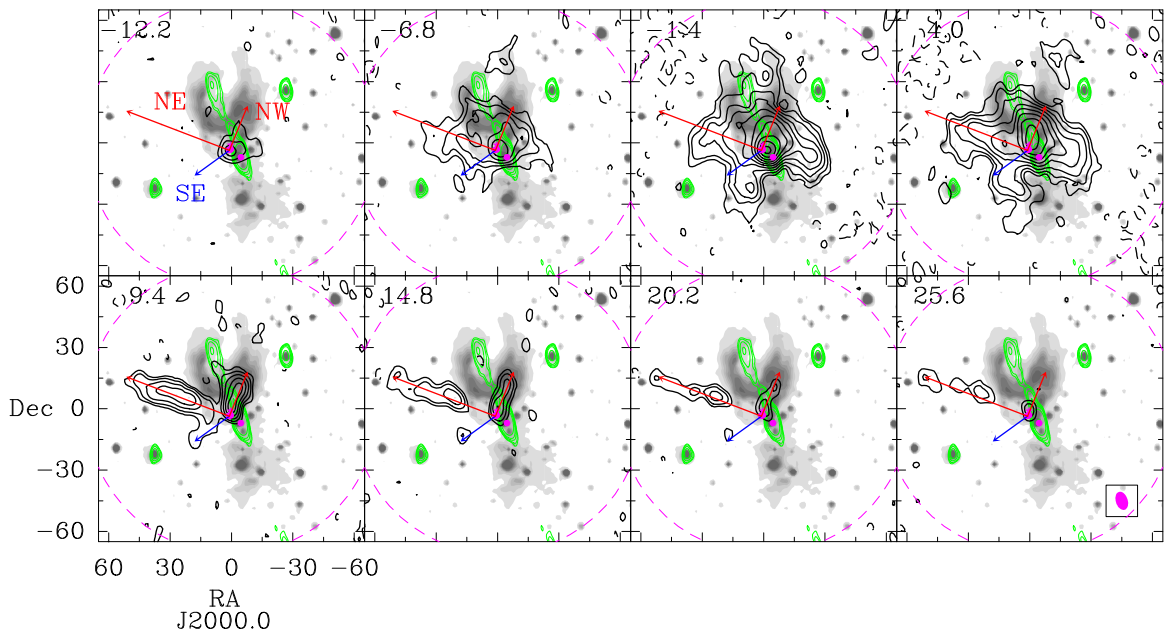}
\caption{HCN(1-0) emission velocity map (black contours) overlapped with a VLA HH~80/81/80N radio jet image (green contours) at 6~cm and a 2MASS K-band infrared image (grey scale). HCN(1-0) contours are -3, 3, 5, 8, 12, 17, 23, 30, 38, 47 and 57 $\times$ 30\mjy, the rms noise of the image. Symbols are the same that those of Fig. \ref{moms0}.}
\label{hcn}
\end{figure*}

\begin{figure*}
\epsscale{1}
\plotone{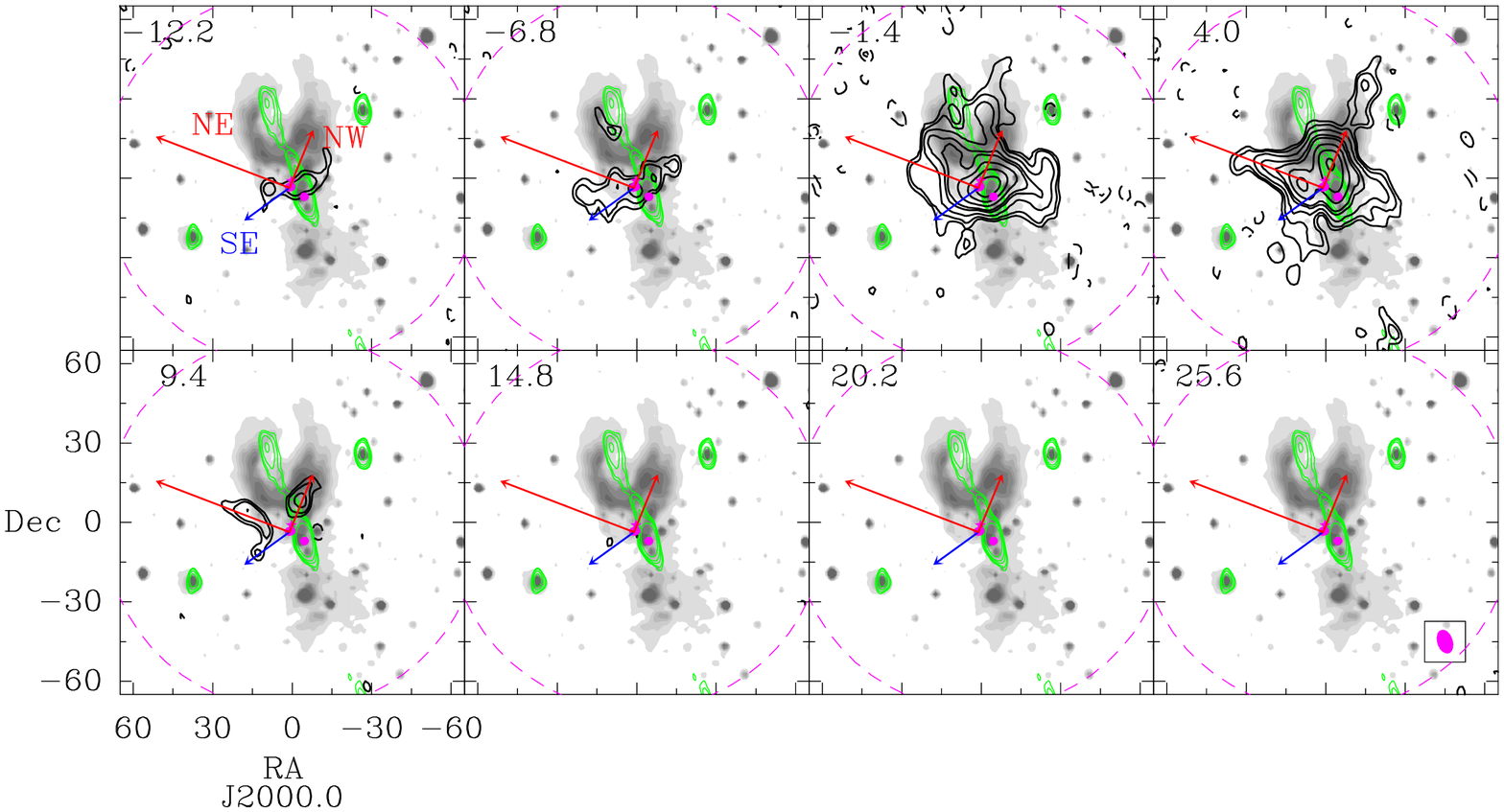}
\caption{HCO$^+$(1-0) emission velocity map (black contours) overlapped with a VLA HH~80/81/80N radio jet image (green contours) at 6~cm and a 2MASS K-band infrared image (grey scale). HCO$^+$(1-0) contours are -3, 3, 4, 6,  9, 11, 16, 22 and 29 $\times$ 30\mjy, the rms noise of the image. Symbols are the same that those of Fig. \ref{moms0}.}
\label{hco}
\end{figure*}

\begin{figure*}
\epsscale{1}
\plotone{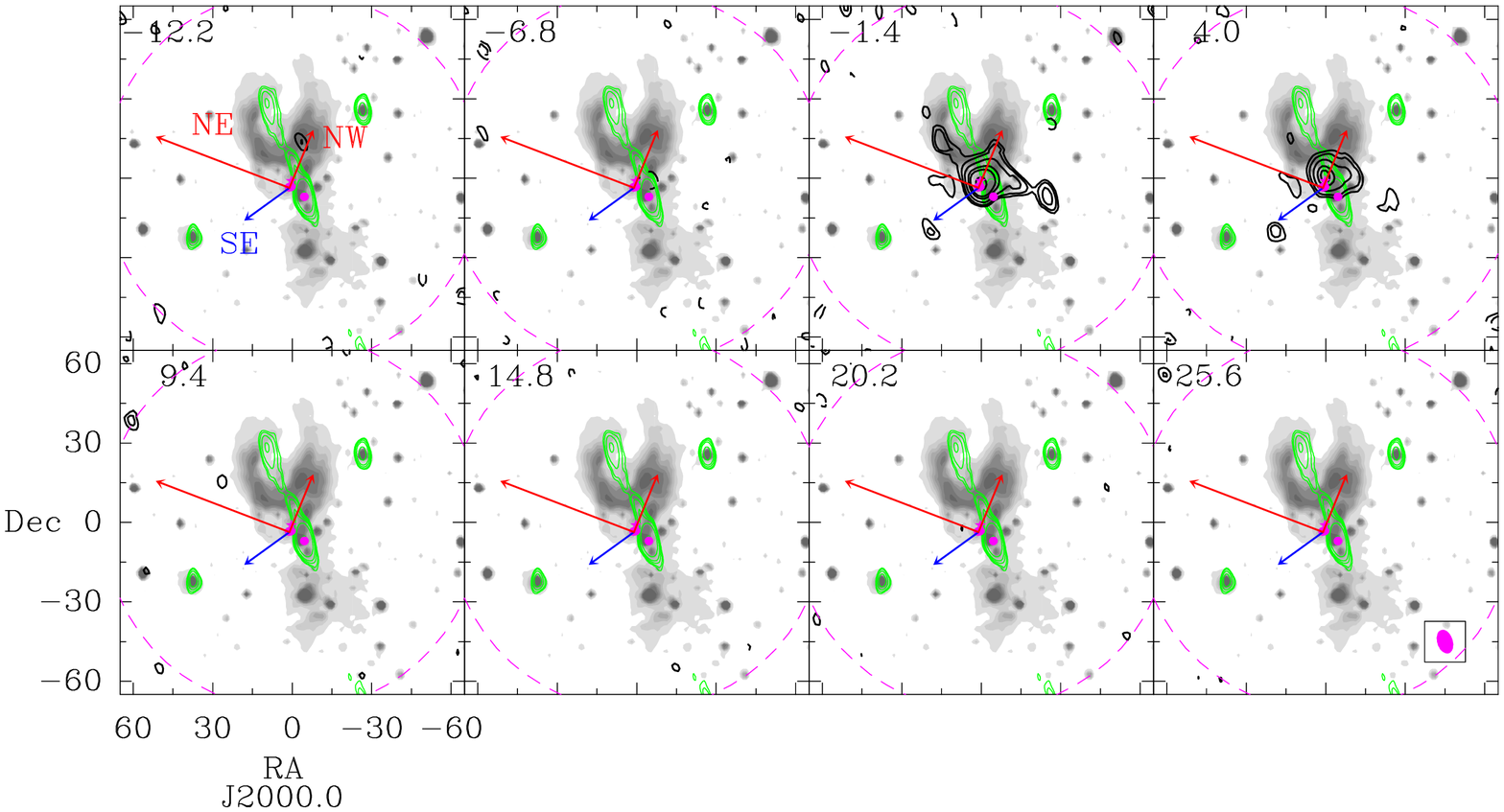}
\caption{HC$_3$N(10-9) emission velocity map (black contours) overlapped with a VLA HH~80/81/80N radio jet image (green contours) at 6~cm and a 2MASS K-band infrared image (grey scale). HC$_3$N(10-9) contours are -3, 3, 4, 6, 9, 11 and 16 $\times$ 35\mjy, the rms noise of the image. Symbols are the same that those of Fig. \ref{moms0}.}
\label{hc3n}
\end{figure*}
 
\section{Results}
\subsection{Continuum emission at 3.5~mm}
Using CARMA we detected a resolved 3.5~mm source composed of two components (Fig. \ref{cont}). The spatial distribution of the continuum emission resembles very well that of the continuum at 1.4~mm shown by \citealt{2011FernandezLopez1}. We applied a two-component Gaussian model to fit the continuum data, which left no residuals over a 3-$\sigma$ level. From this fit, the main southwestern component is spatially coincident with the exciting source of the thermal jet, MM1, and has an integrated flux of 94\mjy. The component to the northeast coincides with the position of MM2 and has an integrated flux of 36\mjy. This is the first time that a measurement of MM2 can be obtained at 3~mm, and it is in good agreement with the flux density expected at this wavelength from the estimated spectral index for MM2, using previous millimeter and submillimeter observations (\citealt{2011FernandezLopez1}). Although no continuum emission from the molecular core (MC) detected by \cite{2009Qiu} was needed to explain the whole continuum emission, the angular resolution of the present CARMA observations does not allow us to rule out the possibility that the molecular core contributes to the observed dust emission.

The total flux measured on the field of view is $145\pm15$\mjy, which is also consistent with the total flux reported by \citealt{2003Gomez}. However, they proposed that all the emission is associated with MM1, while we find that a large fraction of the emission is associated with MM1 and the rest of the emission comes from MM2.

\subsection{Molecular emission}
\subsubsection{Detection of outflows}
CARMA observations of the classical outflow tracers SiO(2-1), HCN(1-0) and HCO$^+$(1-0) were aimed to studying the emission from the outflow associated with MM1 and its powerful radio jet (\citealt{2001Ridge,2004Benedettini}). We do not detect any emission associated with this radio jet in the velocity range (-211,+189)\kms covered by the observations, although part of the HCN and HCO$^+$ low-velocity emission could be associated with material encasing the collimated radio jet. In what follows we adopt $v_{lsr}=11.8$\kms as the cloud velocity (\citealt{2011FernandezLopez2}). The CARMA observations do not show molecular emission from the southeast monopolar outflow associated with MM2 and previously reported by \cite{2009Qiu}, but they show two other monopolar outflows, possibly originated from MM2 and/or MC (Fig. \ref{moms0}).

The SiO(2-1) arises only at redshifted velocities (from +4 to +38\kms with respect to the cloud velocity\footnote{All the velocities in this work are given with respect to the cloud velocity, adopted as  $v_{lsr}=11.8$\kms.}) and from two different spots: a well collimated lobe running in the northeast direction, east of MM2, and a low collimated lobe running in the northwest direction, north of MM2 and apparently coinciding with the infrared reflection nebula (e.g., \citealt{1992Aspin}) seen in grey scale in Fig \ref{sio} (which only shows the channels with the main emission from -12 to +26\kms). 

Fig. \ref{hcn} shows the HCN(1-0) velocity cube. The spectral resolution of these observations does not resolve the hyperfine structure of this line, since the two strongest transitions are within 5\kms. The redshifted HCN(1-0) emission mainly coincides with the same two outflows seen in SiO(2-1), but with lower velocities (near the cloud velocity), and the emission is more spread out, nearly matching the spatial distribution of the infrared reflection nebula and following the radio jet path. At these velocities, the HCN(1-0) also shows elongated emission due southwest (see also Fig. \ref{moms0}). 

HCO$^+$(1-0) redshifted emission is weaker, but also coinciding with the two SiO(2-1) outflows (Fig. \ref{hco}). The low velocity emission from this molecular line also appears extended and mostly associated with the infrared reflection nebula and possibly the radio jet path.

In addition to CARMA observations, we present here new high-angular resolution SMA observations showing CO(3-2) emission from the southeast outflow. Fig. \ref{co32} shows the CO(3-2) SMA images in three panels showing three different velocity regimes: low (LV), medium (MV) and high velocity (HV). The southeast outflow appears in all of them, comprised of blue and redshifted emission at LV and MV, and mostly blueshifted emission at HV. The HV panel, the map with highest angular resolution, clearly shows that the origin of the southeast outflow is MM2(E). Inspecting that image, it is evident that this outflow is wiggling, being ejected due east at the origin, and then turning southeast (about $\sim30\degr$ change in position angle). Such behavior can be produced by precessing motion of the source (e.g., Raga et al. 2009). The middle panel shows a possible additional east turn in the blueshifted emission far away from MM2(E).  Finally, the LV and MV maps also show the redshifted northwest lobe and hints of the northeast lobe, but the angular resolution of these maps do not allow identification of their origin. 

From now on we will call the two high velocity SiO lobes, NE (Northeast, P.A.$=71\degr$) and NW (Northwest, P.A.$=-22\degr$) outflows. The southeast outflow (P.A.$=126\degr$) will be designated as the SE outflow. 

\begin{deluxetable*}{lcccccc}[h]
\tablewidth{0pc}
\tablecolumns{7}
\tabletypesize{\scriptsize}
\tablecaption{Outflow parameters}
\tablehead{
\colhead{} & \colhead{NE outflow} & \colhead{NW outflow} & \colhead{SE outflow\tablenotemark{(a)}} & \colhead{Low-mass Class 0\tablenotemark{(b)}} & \colhead{$L<10^3$\lsun\tablenotemark{(c)}} & \colhead{$L\in(10^3,5\times10^3)\tablenotemark{(d)}$\lsun}
}
\startdata
N(SiO) ($10^{13}$ cm$^{-2}$) & 1.2$\pm0.2$      & 1.1$\pm0.2$  & \nodata & \nodata & \nodata & \nodata \\
\hline
\\
Position angle (\degr)                      & 71$\pm$3         & -22$\pm$3 & 126 & \nodata & \nodata & \nodata \\
$\lambda$ (pc)                               & 0.54$\pm$0.03 &  0.13$\pm$0.03 & 0.2 & \nodata & \nodata & \nodata \\
v$_{max}$ (\kms)                          & 38$\pm$3        &  27$\pm$3 & 100 & \nodata & \nodata & \nodata \\
t$_{dyn}$ ($10^3$ yr)                   & 18$\pm$2        & 5$\pm$2 & 2.2 & \nodata & 10 & 38\\
\hline 
\\
M (\msun)                                                 & 1.60$\pm$0.08      & 0.74$\pm$0.05 & 0.22 & $10^{-3}-0.4$ & 2 & 10\\
$\dot{M}$ (10$^{-5}$ \msun~yr$^{-1}$) & 9$\pm$1            & 14$\pm$7 & 10 & \nodata & 21 & 37 \\
P (\msun~\kms)                     		        & 24$\pm$6          & 9$\pm$3 & 4.9 & $\sim0.02$  & 16 & 54 \\
$\dot{P}$ (10$^{-3}$\msun~\kms~yr$^{-1}$) & 1.3$\pm$0.5 & 1.9$\pm$1 & 2.2 & \nodata & 1.7 & 1.9\\
E ($10^{45}$ erg)                                    & 5.0$\pm$0.6            & 1.5$\pm$0.2 & \nodata & $10^{-4}-10^{-3}$ & 2 & 4\\
L$_{mech}$ (\lsun) 			        & 2.3$\pm$0.5      & 2$\pm$1 & \nodata & \nodata & 2 & 1
\enddata 
\tablecomments{Outflow parameters from top to bottom: SiO column density (N(SiO)), position angle, length ($\lambda$), maximum velocity (v$_{max}$), dynamical time (t$_{dyn}$), mass (M), mass injection rate ($\dot{M}$), momentum (P), momentum rate ($\dot{P}$), energy (E),  and mechanical luminosity (L$_{mech}$). The uncertainties are just statistical and the inclination with respect to the plane of the sky has not been taken into account.}
\tablenotetext{(a)}{Data extracted from \cite{2009Qiu}.}
\tablenotetext{(b)}{Values were extracted from low-mass Class 0 sources observed in several works, see text.}
\tablenotetext{(c)}{Data relative to outflows from high-mass protostars with luminosities below $10^3$\lsun were averaged from observations by \cite{2005Zhang}.}
\tablenotetext{(d)}{Data relative to outflows from high-mass protostars with luminosities between $10^3$\lsun and $5\times10^3$\lsun were averaged from observations by \cite{2005Zhang}.}
\label{outflows}
\end{deluxetable*}

\subsubsection{Gas tracing the reflection nebula}

Figures \ref{moms0}, \ref{sio}, \ref{hcn} and \ref{hco} allow comparisons between the observed molecular emission and the 2MASS K-band emission from the infrared reflection nebula and the 6~cm VLA radio continuum emission from the radio jet launched from MM1.
The HCN(1-0) and HCO$^+$(1-0) lines are tracing the gas from the molecular outflows, but also other kind of structures. We have also detected emission from the HC$_3$N(10-9) transition, which is weaker than the other detected lines and at velocities close to systemic (Fig. \ref{hc3n}).
  
The 2MASS K-band image shows the well-known bipolar reflection nebula (\citealt{1991Aspin,1992Aspin}). This nebula wraps the radio jet path. Toward the position of MM1, the nebula becomes narrower and splits into two U--shaped lobes. The north lobe, with an intricate structure, matches quite well most of the HCN(1-0), HCO$^+$(1-0) and HC$_3$N(10-9) emission between -12 and +4\kms. 
The blueshifted molecular line emission (-12 to -7\kms) traces the basis of the reflexion nebula and spreads out due north, covering the whole northern lobe at velocities in the range -1\kms and +4\kms. At these velocities, part of the HCN(1-0) and HCO$^+$(1-0) emission follows the radio jet trajectory towards the north of the MC position.
The southern lobe has fainter K-band emission than the northern lobe and it has no molecular emission associated with it. The lack of strong molecular line emission in that area of the nebula is probably due to the lack of dense molecular gas, as shown in the C$^{17}$O(2-1) emission map by (\citealt{2011FernandezLopez2}).

\section{Analysis of the outflow properties from SiO emission}
SiO is a good tracer of outflows (e.g. \citealt{2010JimenezSerra,2011LopezSepulcre} and references therein). Its abundance is dramatically increased in shocks (e.g., \citealt{1997Schilke,1996PineauDesForets,2008Gusdorf2,2008Gusdorf1, 2009Guillet}) and has been used as a tool to map the innermost part of outflows \citep{1992MartinPintado,2009SantiagoGarcia}. In addition, it does suffer minimal contamination from quiescent and infalling envelopes and can not be masked by blended hyperfine components as in the case of the HCN(1-0) line. Thus, we choose this molecular outflow tracer to derive the characteristics of the NE and NW outflows.

SiO(2-1) has been previously detected toward \iras with the IRAM 30~m by \cite{1997Acord}. They reported an integrated flux of $1.9\pm0.4$~K~\kms, while here we have measured $2.5\pm0.05$~K~\kms. Applying a $27\arcsec$ beam dilution (the angular size of the IRAM 30~m beam), the CARMA measurement becomes 1.9~K~\kms. This implies that CARMA is recovering the same flux as the IRAM 30~m for this transition and hence, we can use it to estimate some physical parameters of the outflow. 

In the first place, we estimate the column density using the following equation (derived from the expressions contained in the appendix of \citealt{2010Frau}), written in convenient units:
$$N_{H_2} =2.04\times10^{20}\; \chi(SiO) \;\frac{Q(T_{ex})\; e^{E_u/T_{ex}}}{\Omega_s\; \nu^3 \; S\mu^2}\, \int{S_{\nu}\,dv}\quad,$$
where $\chi(SiO)$ is the abundance of SiO relative to H$_2$, $Q(T_{ex})$ is the partition function, $E_u$ and $T_{ex}$ are the upper energy level and the excitation temperature, both in K, $\Omega$ is the angular size of the outflow in square arcseconds, $\nu$ is the rest frequency of the transition in GHz, $S\mu^2$ is the product of the intrinsic line strength in erg~cm$^3$~D$^{-2}$ and the squared dipole momentum in D$^2$.  Following the NH$_3$ observations of \cite{2003Gomez}, we adopted an excitation temperature of 30~K. $E_u$, $Q(T_{ex})$ and $S\mu^2=19.2~erg~cm^3$ were extracted from the CDMS catalogue (Muller et al. 2005). The term $\int{S_{\nu}\,dv}$ is the measured integrated line emission in Jy~beam$^{-1}$~\kms. 

We analyzed the two SiO outflows separately (NE and NW outflows), finding average column densities of $1.1\pm0.2\times10^{13}$\cmd and  $1.2\pm0.2\times10^{13}$\cmd respectively. We used these values to derive the mass ($M=N_{H_2}\mu_g m_{H_2}\Omega_s$) and other properties of the outflow, assuming a mean gas atomic weight $\mu_g=1.36$. The results are shown in Table \ref{outflows}. 

We used an SiO abundance of $\chi(SiO)=5\times10^{-9}$. This value is in good agreement with the $\chi(SiO)$ found in outflows driven by other intermediate/high-mass protostars. For instance, in a recent paper, \cite{2013SanchezMonge} found the $\chi(SiO)$ in 14 high-mass protostars outflows range between $10^{-9}$ and $10^{-8}$ (similar values were also encountered in \citealt{2001Hatchell, 2007Qiu,2013Codella}).
It is important to mention that the uncertainty in the SiO abundance is probably one of the main sources of error for our estimates (e.g., \citealt{2009Qiu2}). One order of magnitude in $\chi(SiO)$ translates into one order of magnitude in most of the outflow properties given in Table \ref{outflows} (M, $\dot{M}$, P, $\dot{P}$, E and L$_{mech}$). Most of the outflow properties in this table were derived following the approach of \cite{2007Palau}.
In order to derive the mass and momentum rates, together with the mechanical luminosity, we need to know the dynamical timescale of the outflows, which can be determined as $t_{dyn}=\lambda/v_{max}$, where $\lambda$ is the outflow length. No correction for the inclination was included. Hence, we estimate the dynamical time of the NE and NW outflows as 18000~yr and 5000~yr, and their masses as 1.60 and 0.74\msun, respectively (Table \ref{outflows}). In addition, it is worth to note that most of the mass (about 80\%) of the NE outflow is concentrated in its brightest condensation (see Fig. \ref{moms0}).

\section{Discussion}
\subsection{Asymmetric outflows and their possible origin}

Recent work has provided evidence supporting MM1 being a 11-15\msun high-mass protostar, probably still accreting material from a 4\msun rotating disk, and MM2 being a massive dusty core containing at least one high-mass protostar, MM2(E), in a less evolved stage than MM1 (\citealt{2011FernandezLopez2}). MM2 also contains another core, MM2(W), thought to be in a still earlier evolutionary stage. In addition to these well known protostars, at present it is controversial whether MC harbors another protostar, since neither a thermal radio continuum nor a dust continuum emission has been detected in this molecular core. Hence in the area MM2-MC we can count two or three protostars that could be associated with the molecular outflows reported in this work.

\iras has protostars surrounded by accretion disks and/or envelopes, and these protostars are associated with molecular outflows and jets, resembling low-mass star-forming systems. The most apparent case would be that of MM1, with one of the largest bipolar radio jets associated with a protostar. Being a scaled up version of the low-mass star formation scenario implies more energetic outflows (as we see in MM1 and MM2), which means larger accretion rates, but it also implies large amounts of momentum impinged on the surrounding environment, affecting occasionally the nearby protostellar neighbors and their outflows. In this case, \iras is a region with multiple outflows, mostly seen monopolar. Our observations show no well opposed counterlobes for the NW, NE and SE outflows. That could be explained if the counterlobes are passing through the cavity excavated by the radio jet or through regions of low molecular abundance. Another explanation for the asymmetry of outflows, perhaps more feasible in a high-mass star-forming scenario with a number of protostars and powerful outflows, is deflection after hitting a dense clump of gas and dust (e.g., \citealt{2002Raga}).

Adding a little bit more to the complexity of the region, the SE outflow is comprised of two very different kinematical components. Fig. \ref{co32} shows high velocity emission at two well separated velocities laying almost at the same spatial projected path. A blueshifted and redshifted spatial overlap has been usually interpreted as an outflow laying close to the plane of the sky. However, the SE outflow has large radial blueshifted and redshifted velocities ($\sim\pm50$\kms). 
One possibility is that the high velocity emission of the SE outflow comes from two molecular outflows, one redshifted and the other one blueshifted, both originated around the MM2 position.
A second possibility is that the SE outflow is precessing with an angle $\alpha=15\degr$ (see Fig. 2 in \citealt{2009Raga}). We derived this angle by taking half the observed wiggling angle of the outflow axis in projection (from P.A.$\simeq95\degr$ at the origin of the outflow to P.A.$\simeq126\degr$ away from it). Hence, the absolute velocity of the ejecta should be $\sim200$\kms, which seems to be a reasonable value (see e.g., \citealt{2009Bally}). A smaller $\alpha$ angle would imply a higher outflow velocity.
From the central panel of Fig. \ref{co32}, we roughly estimate the period of the wiggles of the SE outflow as $\lambda\simeq 16\arcsec$ (27000~AU). We have assumed that the SE outflow completes one precessing period between the position of MM2(E) and the end of the CO(3-2) blueshifted emission in our map. Thus, the precession period is $\tau_p=\lambda / (v_j~\cos{\alpha}) \simeq 660~yr$. If the precession is caused by the tidal interaction between the disk of a protostar in MM2(E) and a companion protostar in a non-coplanar orbit (e.g. \citealt{1999Terquem1,2009Montgomery}) then it is possible to obtain some information about the binary system. We use an equivalent form of equation (37) from \cite{2009Montgomery} for circular precessing Keplerian disks. This equation relates the angular velocity at the disk edge ($\omega_d$), the Keplerian orbital angular velocity of the companion around the primary protostar ($\omega_o$) and the retrograde precession rate of the disk and the outflow ($\omega_p$): 
$$\omega_p=-\frac{15}{32}\frac{\omega_o^2}{\omega_d}\cos{\alpha}\quad,$$
where $\alpha$ is the inclination of the orbit of the companion with respect to the plane of the disk (or obliquity angle), and this angle is the same as the angle of the outflow precession (i.e. the angle between the outflow axis and the line of maximum deviation of the outflow from this axis; \citealt{1999Terquem2}). Using this expression and adopting reasonable values for the mass of the primary protostar and the radius of its disk we can constrain the orbital period and the radius of the companion protostar. \cite{1995Gomez} assigned a B4 ZAMS spectral type to MM2(E) based on its flux at 3.5~cm. A B4 spectral type protostar has about 6-7\msun (Table 5 in \citealt{1998Molinari}). On the other side, the dust emission of MM2(E) has a radius no larger than 300~AU (\citealt{2011FernandezLopez1}) and we consider 50~AU as a reasonable lower limit for the disk radius. With all of this we derived an orbital period between 200~yr and 800~yr and an orbital radius between 35~$M_2^{1/3}$~AU and 86~$M_2^{1/3}$~AU for the putative MM2(E) binary system, being $M_2$ the mass of the secondary protostar expressed in solar masses.  

The HV panel of Fig. \ref{co32} clearly shows for the first time the precise origin of the SE-blueshifted outflow: MM2(E). Its corresponding redshifted counterlobe appears truncated between MM2(E) and MM2(W). However, the CO(3-2) LV and MV panels show that the redshifted NW outflow reaches the position of MM2 (see also HCN(1-0) emission in Fig. \ref{hcn}). Then, it is possible that the SE outflow counterpart is the NW outflow. This resolves the monopolar nature of both the NW and SE outflows and explains the prominent redshifted wing spectral line profiles of H$_2$CO and SO transitions previously observed by \cite{2011FernandezLopez2} at the position of MC, as being produced by a receding outflow from MM2(E) colliding with a dense cloudlet at the position of MC. The change in the SE-NW outflow direction could be explained by a deflection due north of the NW outflow. The cause of the deflection could be a direct impact against MM2(W), or the action of the powerful HH~80/81/80N wind over the NW lobe. 

The NE outflow has a monopolar structure at first sight too (Figs. \ref{sio}, \ref{moms0} and \ref{hcn}). Its origin cannot be well determined due to angular resolution constraints, but it is also in the MM2-MC area.
Figs. \ref{moms0}, \ref{hcn} and \ref{hco} (channels at -1 and +4\kms) show some signs that a low-velocity counterlobe may exist with a position angle of -112$\degr$, almost opposite to the NE outflow. It would spread out 0.20-0.25~pc from the MM2-MC position.
 
\subsection{Evolutionary stage of outflows and protostars}
As stated in several works, SiO is a commonly used molecular tracer of shocked gas in outflows from low-mass protostars (e.g., \citealt{2006Hirano,2008Lee,2009SantiagoGarcia}), but SiO is also found in outflows from high-mass protostars (e.g., \citealt{1999Cesaroni,2001Hatchell,2009Qiu2,2004Beuther,2007Zhang2,2007Zhang1,2011LopezSepulcre,2012Zapata1,2013Leurini}). However, with the present CARMA observations we have not detected SiO(2-1) emission associated with the HH~80/81/80N radio jet, nor with the SE outflow along the $\sim400$\kms of the CARMA SiO(2-1) window bandwidth. The HH~80/81/80N and the SE outflows have not been observed either in the other molecular transitions of this study, HCO$^+$(1-0) and HCN(1-0), also known to be good outflow tracers. 
If anything, some HCO$^+$ and HCN emission may come from gas pushed away by the collimated and high-velocity jets or may be due to low-velocity winds.
Both outflows have been observed in CO lines at the velocities sampled by the CARMA observations, though. Then, what is producing the different chemistry in the outflows of the region? Why is the SiO(2-1) not detected in the SE outflow nor the radio jet, while the other two outflows NE and NW are? Furthermore, why is only one lobe of each the NE and NW outflows detected? There are other cases in the literature where a similar behavior is observed in CO and SiO (e.g., \citealt{2007Zhang2,2007Zhang1,2008Reid,2012Zapata1,2013Codella}). 

It has been proposed that the SiO abundance can decrease with the age of the outflow (\citealt{1999Codella,2006Miettinen,2010Sakai,2011LopezSepulcre}), which could explain the differences of SiO emission from the outflows of the same region. This hypothesis implies that during the early stages, the gas surrounding the protostar is denser and rich in grains, producing stronger shocks between the outflow and  the ambient material, and thus producing an abundant release of SiO molecules. After that, in more evolved stages, the outflow digs a large cavity close to the protostar and thus the shocks are weaker, and grains are rarer. The hypothesis has further support in the shorter SiO depletion timescale (before it freezes out onto the dust grains) with respect to the typical outflow's timescale (some $10^4$~yr), together with its disappearance from the gas phase favoring the creation of SiO$_2$ (\citealt{1996PineauDesForets,2004Gibb}) and could describe well the HH~80/81/80N jet case, since it produced a cavity probably devoid of dust grains.
Now, we can compare the timescale of the outflows in the central region of \iras. The radio jet HH~80/81/80N is 10$^6$~yr (\citealt{2004Benedettini}), the SE outflow has an age of about $2\times10^3$ years (\citealt{2009Qiu}) and the NE and NW outflows have $2\times10^4$ and $5\times10^3$ years. Therefore, except for the SE outflow, the outflow timescales would be in good agreement with SiO decreasing its abundance with time. Actually, as indicated before, the case for the SE outflow is more complex. It has high velocity gas and it is apparently precessing. That can therefore indicate a  larger outflow path. In addition CO(3-2) observations are constrained by the SMA primary beam, implying that the outflow could be larger than observed and thereby older. In any case, if the SE outflow is the counterlobe of the NW one (see \S5.1), then a different explanation must be found to account for the chemical differences between these two outflow lobes.
 
We can also compare the characteristics of the NE and NW outflows with those of outflows ejected by (i) high-mass protostars and (ii) low-mass Class 0 protostars, in order to put additional constrains on the  ejecting sources. \cite{2005Zhang} made an outflow survey toward high-mass star-forming regions using CO single-dish observations. After inspecting this work we summarize the properties belonging to outflows from $L<10^3$\lsun protostars and $L\in(10^3,5\times10^3)$\lsun protostars in Table \ref{outflows}. This Table shows in addition, information on outflows from low-mass Class 0 protostars as well, gathered up from several sources (\citealt{2004Arce,2005Arce,2006Kwon,2011Davidson}).
The characteristics of the NE and NW outflows in \iras (as well as those of the SE outflow) are similar to outflows from high-mass protostars, being the NE outflow more energetic and with higher momentum than the NW and SE outflows. On the contrary, the properties of outflows from low-mass Class 0 protostars, although similar in length and dynamical time, have in general lower mass, and overall kinetic energy, about four orders of magnitude lower. All of this indicates that the NE, NW and SE outflows in \iras could be associated with intermediate or high-mass protostars in a very early evolutive stage (massive class 0 protostars). 
Therefore, given the powerful outflowing activity from MM2, the protostars would be undergoing a powerful accretion process in which the gas from the dusty envelope (about 11\msun) is probably falling directly onto the protostars.
This kind of objects is very rare.  Maybe the closest case is that of Cepheus E (\citealt{2003Smith}). The outflow from this intermediate-mass protostar, which is surrounded by a massive $\sim25$\msun envelope, is very young (t$_{dyn}\sim1\times10^3$~yr), with a mass and an energy (M$\sim0.3$\msun, E$\sim5\times10^{45}$~ergs) resembling those obtained for the NE, NW and SE outflows.

\section{Conclusions}
We have carried out CARMA low-angular resolution observations at 3.5~mm and SMA high-angular resolution observations at 870~$\mu$m toward the massive star-forming region \iras. We have also included the analysis of SMA low-angular resolution archive data analysis of the CO(3-2) line. The analysis of several molecular lines, all of which are good outflow tracers, resulted in the physical characterization of two previously not well detected outflows (NE and NW outflows) and the clear identification of the driving source of a third outflow (SE outflow). The main results of this work are as follows:

\begin{itemize}
\item We observed three apparently monopolar or asymmetric outflows in \iras. The NE and NW outflows were detected in most of the observed molecular lines (SiO, HCN, HCO$^+$ and CO), while the SE outflow was only clearly detected in CO. The outflow associated with HH~80/81/80N was undetected. At most, it could explain some HCN and HCO$^+$ low-velocity emission associated with the infrared reflection nebula, which could be produced by dragged gas or a wide open angle low-velocity wind from MM1.
\item The NE and NW outflows have their origins close to MM2. 
\item We have estimated the physical properties of the NE and NW outflows from their SiO emission. They have similar characteristics to those found in molecular outflows from massive protostars, being the NE outflow more massive and energetic than the NW and SE outflows.
\item SMA high-angular resolution CO(3-2) observations \emph{have identified} the driving source of the SE outflow: MM2(E). These observations provide evidence of precession along this outflow, which show a change of about 30$\degr$ in the position angle and a period of 660~yr. If the precession of the SE outflow is caused by the misalignment between the plane of the disk and the orbit of a binary companion, then the orbital period of the binary system is 200-800~yr and the orbital radius is 35-86~$M_2^{1/3}$~AU.
\item We discuss the monopolar or asymmetric appearance of all three outflows. We provide evidence that the SE and NW outflows are linked and that precession and a possible deflection are the causes of the asymmetry of the outflow. In addition, the NE outflow could have a smaller and slower southwest counterlobe, maybe associated with elongated HCN and HCO$^+$ emission. 
\item Finally, we discuss that the SiO outflow content in \iras could be related to outflow age. This would explain the SiO non-detection of the radio jet HH~80/81/80N.  
\end{itemize}

\acknowledgments
The authors want to bring a special reminder of our good fellow Yolanda G\'omez, who helped in the very beginning of this work with her contagious optimism.  

We thank all members of the CARMA and SMA staff that made these observations possible. We thank Pau Frau for helping with the SMA observations.
MFL acknowledges financial support from University of Illinois and thanks John Carpenter and Melvin Wright for their patience with CARMA explanations. MFL also thanks the hospitality of the Instituto de Astronom\'{\i}a (UNAM), M\'exico D.F., and of the CRyA, Morelia.
JMG are supported by the Spanish MICINN AYA2011-30228-C03-02 and the Catalan AGAUR 2009SGR1172 grants.
SC acknowledges support from CONACyT grants 60581 and 168251. LAZ acknowledges support from CONACyT.

Support for CARMA construction was derived from the Gordon and Betty Moore Foundation, the Kenneth T. and Eileen L. Norris Foundation, the James S. McDonnell Foundation, the Associates of the California Institute of Technology, the University of Chicago, the states of Illinois, California, and Maryland, and the National Science Foundation. Ongoing CARMA development and operations are supported by the National Science Foundation under a cooperative agreement, and by the CARMA partner universities.

The Submillimeter Array is a joint project between the Smithsonian Astrophysical Observatory and the Academia Sinica Institute of Astronomy and Astrophysics, and is funded by the Smithsonian Institution and the Academia Sinica.

Facilities: \facility{CARMA, SMA}

\bibliography{biblio}

\newpage

\end{document}